\documentstyle[multicol,psfig,prl,aps]{revtex}
\begin{document}
\draft

\title{
Exact Solution of the Statistical Multifragmentation Model\\
and Liquid-Gas Phase Transition in Nuclear Matter
}

\author{\bf K.A. Bugaev$^{1,2}$, M.I. Gorenstein$^{1,2}$,
I.N. Mishustin$^{1,3,4}$ and W. Greiner$^{1}$
}
\address{
$^1$ Institut f\"ur Theoretische Physik,
Universit\"at Frankfurt, Germany\\
$^2$ Bogolyubov Institute for Theoretical Physics,
Kyiv, Ukraine\\
$^3$ The Kurchatov Institute, Russian Research Center,
Moscow, Russia\\
$^4$ The Niels Bohr Institute, University of Copenhagen, Denmark
}
\date{\today}

\maketitle

\begin{abstract}
An exact analytical solution of the 
statistical  multifragmentation model is found in thermodynamic limit.
The system of nuclear fragments exhibits a 1-st order 
liquid-gas phase transition.
The peculiar thermodynamic properties of the model near the boundary
between the mixed phase
and the pure gaseous phase are studied.
\end{abstract}

\vspace*{0.5cm}

\noindent
\hspace*{2.cm}\begin{minipage}[t]{14.cm}
\end{minipage}

\vspace*{0.3cm}


\begin{multicols}{2}

Nuclear multifragmentation is one of 
the most interesting and widely discussed phenomena in
intermediate energy nuclear reactions.
The statistical multifragmentation model (SMM) 
(see \cite{Bo:95,Gr:97} and references therein)
was recently applied
to study the relationship
of this phenomenon to the liquid-gas phase transition
in nuclear matter
\cite{Ch:95,Gu:98,Gu:99}. Numerical calculations
within the canonical ensemble exhibited
many intriguing peculiarities of the finite multifragment
systems. 
However, the investigation of the system's
behavior in the thermodynamic limit is still missing.
Therefore, there is no rigorous proof of the phase
transition existence, and the phase diagram structure
of the SMM is unknown yet. 
In the present paper an exact analytical solution
of the SMM is found within the grand canonical ensemble. 
The self-consistent treatment
of the excluded volume effects
is an important part of our study. 

The system states in the SMM are specified by the multiplicity
sets  $\{n_k\}$
($n_k=0,1,2,...$) of $k$-nucleon fragments.
The partition function of a single fragment with $k$ nucleons is
\cite{Bo:95}:
$
\omega_k =V\left(m T k/2\pi\right)^{3/2}~z_k~
$,
where $k=1,2,...,A$ ($A$ is the total number of nucleons
in the system), $V$ and $T$ are, respectively, the  volume
and the temperature of the system,
$m$ is the
nucleon mass.
The first two factors in $\omega_k$ originate
from the
non-relativistic thermal motion 
and the last factor,
 $z_k$, represents the intrinsic partition function of the
$k$-fragment.
For \mbox{$k=1$} (nucleon) we take $z_1=4$ 
(4 internal spin-isospin states) 
and for fragments with $k>1$ we use the expression motivated by the
liquid drop model (see details in \mbox{Ref. \cite{Bo:95}):} 
$
z_k=\exp(-f_k/T),
$ with fragment free energy 
\mbox{$
f_k\equiv - [W_{\rm o}+
T^2/\epsilon_{\rm o}]k+\sigma (T) k^{2/3}
$.}
Here $W_{\rm o}=16$~MeV is the bulk binding energy per nucleon,
$T^2/\epsilon_{\rm o}$ is the contribution of 
the excited states taken in the Fermi-gas
approximation ($\epsilon_{\rm o}=16$~MeV) and $\sigma (T)$ is the
surface tension which is parameterized 
in the following form:
$
\sigma (T)=\sigma_{\rm o}
[(T_c^2~-~T^2)/(T_c^2~+~T^2)]^{5/4},
$
with $\sigma_{\rm o}=18$~MeV and $T_c=18$~MeV ($\sigma=0$
at $T \ge T_c$).
The canonical partition function (CPF) of nuclear
fragments in the SMM
has the following form:
\begin{equation} \label{Zc}
Z^{id}_A(V,T)~=~\sum_{\{n_k\}}~\prod_{k=1}^{A}~\frac{\omega_k^{n_k}}{n_k!}~
\delta(A-\sum_k kn_k)~.
\end{equation}
In Eq. (\ref{Zc}) the nuclear fragments are treated as point-like objects.
However, these fragments have non-zero proper volumes and
they should not overlap
in the coordinate space. 
In the excluded volume (Van der
Waals) approximation 
this is achieved
by substituting
the total volume $V$
in Eq. (\ref{Zc}) by the free (available) volume 
$V_f\equiv V-b\sum_k kn_k$, where
$b=1/\rho_{{\rm o}}$
($\rho_{{\rm o}}=0.16$~fm$^{-3}$ is the normal nuclear density).  
Therefore, the corrected CPF becomes:
$
Z_A(V,T)=Z^{id}_A(V-bA,T)
$.
The SMM defined by Eq. (\ref{Zc})
was studied numerically in Refs. \cite{Ch:95,Gu:98,Gu:99}.
This is a simplified version of the SMM, e.g. the symmetry and
Coulomb contributions are neglected.
However, its investigation
appears to be of  principal importance
for studies of the liquid-gas phase transition.

The calculation of $Z_A(V,T)$
is difficult because of the constraint $\sum_k kn_k =A$.
This difficulty can be partly avoided by calculating the grand canonical
partition function:
\begin{equation} 
{\cal Z}(V,T,\mu)~\equiv~\sum_{A=0}^{\infty}
\exp\left({\textstyle \frac{\mu A}{T} }\right)
Z_A(V,T)~\Theta (V-bA) \label{Zgc}~, 
\end{equation}
where chemical potential $\mu$ is introduced.
The calculation of ${\cal Z}$  is still rather
difficult. The summation over $\{n_k\}$ sets
in $Z_A$ cannot be performed analytically because of
additional $A$-dependence
in the free volume $V_f$ and the restriction
$V_f>0 $.
The problem can be \mbox{solved \cite{Go:81} }
by introducing the so-called
isobaric partition function (IPF) which is calculated
in a straightforward way:
\begin{eqnarray} \label{Zs}
\hat{\cal Z}(s,T,\mu) & \equiv & \int_0^{\infty}dV~\exp(-sV)
~{\cal Z}(V,T,\mu)  \nonumber\\
& = & \frac{1}{s~-~{\cal F}(s,T,\mu)}~,
\end{eqnarray}
where
\begin{eqnarray}\label{Fs}  
{\cal F}(s,T,\mu)&=&
\left( \frac{mT }{2\pi}\right)^{3/2} 
\left[z_1 \exp\left(\frac{\mu-sbT}{T}\right) \right. 
\nonumber \\
&+& \sum_{k=2}^{\infty} \left.
k^{3/2} \exp\left(
\frac{(\nu - sbT)k -
\sigma k^{2/3}}{T}\right)\right]~, 
\end{eqnarray}
with 
$
\nu \equiv \mu + W_{\rm o}+T^2/\epsilon_{\rm o}
$.
In the thermodynamic limit $V\rightarrow \infty$ the pressure
of the system
is defined by the farthest-right singularity, $s^*(T,\mu)$, of  
the IPF $\hat{\cal Z}(s,T,\mu)$
(see Ref. \cite{Go:98} for details):
\begin{equation}\label{ptmu}
p(T,\mu)~\equiv~ T~\lim_{V\rightarrow \infty}\frac{\ln~{\cal Z}(V,T,\mu)}
{V}~=~T~s^*(T,\mu)~.
\end{equation}
The study of the 
system's behavior in the thermodynamic limit
is therefore reduced to the investigation of
the singularities \mbox{of $\hat{\cal Z}$.}

\vspace*{-0.5cm}

\begin{figure}
\mbox{\psfig{figure=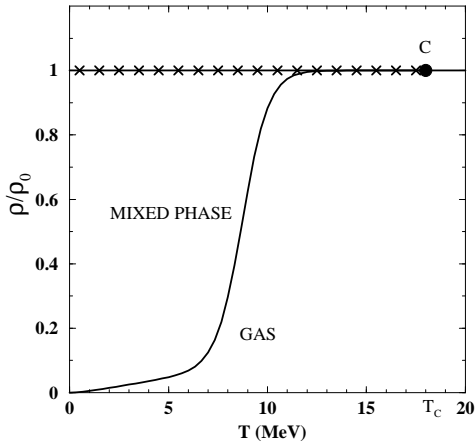,width=98mm}}

\vspace*{-0.5cm}

\caption{\label{fig:one}
Phase diagram in the $(T,\rho)$-plane.
The mixed phase and pure gaseous phase boundary
is shown by the solid line.
The pure liquid phase (shown by crosses) corresponds 
to
the fixed density $\rho = \rho_{\rm o}$.
Point $C$ is the critical point, 
at $T>T_c$ only the pure gaseous phase
exists.
}
\end{figure}

The IPF (\ref{Zs})  has two types of singularities:
\mbox{1) the simple} pole singularity
defined by the equation
\mbox{$s_g(T,\mu)= {\cal F}(s_g,T,\mu)~$;}
2) the singularity  of the function ${\cal F}$ 
 itself at the point $s_l(T,\mu)=\nu/Tb$ where the coefficient 
in linear over $k$ terms of the exponent in Eq. (\ref{Fs}) 
is equal to zero.

The simple pole singularity corresponds to the gaseous phase 
where pressure $p_g\equiv Ts_g$ is determined by the 
transcendental
equation:
$
p_g(T,\mu)=T{\cal F}(p_g/T,T,\mu)
$.
The singularity $s_l(T,\mu)$ of the function ${\cal F}$
defines the liquid pressure:
\mbox{$
p_l(T,\mu)\equiv Ts_l(T,\mu)=
{\nu}/{b}.
$
}
Here the liquid is represented by an infinite fragment with $k=\infty$.

In the region of the $(T,\mu)$-plane where $\nu < bp_g(T,\mu)$ the
gaseous phase
is realized ($p_g > p_l$), while  the liquid phase
dominates at $\nu > b p_g(T,\mu)$. The liquid-gas phase transition
occurs when  the two singularities coincide,
i.e. $s_g(T,\mu)=s_l(T,\mu)$.
As ${\cal F}$ in Eq. (\ref{Fs}) 
is a monotonously decreasing
function of $s$ 
the necessary condition for the phase
transition is that the function
${\cal F}$ is finite in its singular
point $s_l$. 
This condition requires $\sigma(T) >0$ and, therefore, $T<T_c$.
Otherwise, ${\cal F}(s_l,T,\mu)=\infty$ and the system
is always in the gaseous phase as $s_g>s_l$.

At $T<T_c$ the system undergoes a 1-st order phase transition
across the line $\mu^*=\mu^*(T)$ defined by 
the condition of coinciding singularities:
$
s_l=s_g
$
.
The baryonic density $\rho$ in the liquid and gas phases
is given by the following formulae, 
respectively: 
$
 \rho_l  \equiv  
\left(\partial  p_l/\partial \mu\right)_{T}
=1/b
$,
$\rho_g \equiv
\left(\partial  p_g/\partial \mu\right)_{T}=
\rho_{id}/( 1 + b \rho_{id} )
$ ,
where the function $ \rho_{id}$ is the density of point-like
nuclear fragments with shifted, 
$
\mu \rightarrow \mu -bp_g
$,
chemical potential:
\begin{eqnarray}\label{rhoid}
\rho_{id}(T,\mu) &=& \left( \frac{mT }{2\pi}\right)^{3/2} 
\left[z_1 \exp\left(\frac{\mu-bp_g}{T}\right) \right. 
\nonumber \\
&+& \sum_{k=2}^{\infty} \left.
k^{5/2} \exp\left(
\frac{(\nu - bp_g)k -
\sigma k^{2/3}}{T}\right)\right]~.
\end{eqnarray}
A similar expression for $\rho_g$ within the excluded volume model
for the pure nucleon gas was obtained in Ref. \cite{Ri}.
The phase transition line 
$\mu^*(T)$
in the $(T,\mu)$-plane
corresponds to the mixed liquid and gas
states. This line 
is transformed into
the finite mixed-phase region in the $(T,\rho)$-plane
shown in Fig. 1. 
The baryonic density 
in the mixed phase
is a superposition of the liquid and gas baryonic densities:
$
\rho=\lambda\rho_l+(1-\lambda)\rho_g~,
$
where $\lambda$ ($0<\lambda <1$) is a fraction of the system's volume
occupied by the liquid  inside the mixed phase.
Similar linear combinations are also valid for the entropy density $s$
and the energy density $\varepsilon$ 
with $(i=l,g)$
$s_i=\left(\partial p_i/\partial T\right)_{\mu}$,
$\varepsilon_i=T \left(\partial p_i/\partial T \right)_{\mu} +
\mu \left(\partial p_i/\partial \mu\right)_{T} -p_i$.

Inside the mixed phase at constant density $\rho$ the
parameter $\lambda$ has a specific temperature dependence
shown in Fig. 2:
from an approximately
constant value $\rho/\rho_{\rm{o}}$ at small $T$ the function 
$\lambda(T)$ drops to zero in a narrow
vicinity of the boundary separating the  mixed phase and 
the pure gaseous phase.
Such an abrupt decrease of $\lambda(T)$ near this boundary
causes a strong
increase of the energy density as a function of temperature.
This is evident from Fig.~3 which shows the caloric curves at different
baryonic densities. One can clearly see a plateau-like behavior.
As a consequence this  
leads to a sharp peak 
in the specific heat per nucleon at constant density,
$c_{\rho}(T)\equiv (\partial \varepsilon/\partial T)_{\rho}/\rho~$,
shown in Fig. 4.

\vspace*{-1.5cm}

\begin{figure}
\mbox{\psfig{figure=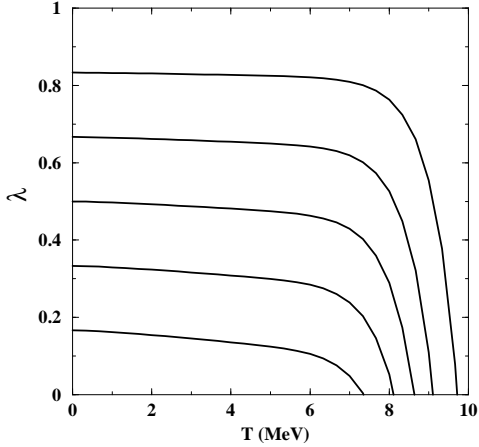,width=98mm}}

\vspace*{-0.5cm}

\caption{\label{fig:two}
Volume fraction $\lambda(T)$ of the liquid
inside the mixed phase is
shown as a function of temperature
for fixed nucleon densities ${\rho}/{\rho_{\rm o}} = 1/6, 1/3, 1/2, 2/3,
5/6$
(from bottom to top).
}
\end{figure}

The peculiar thermodynamic property of the
considered model is
a finite discontinuity of $c_{\rho}(T)$
at the boundary of the mixed and gaseous phases
seen in Fig. 4.
This finite discontinuity
is due to the fact that
$\lambda(T)=0$, but
$(\partial\lambda/\partial T)_{\rho}
\neq 0$ at this boundary inside the mixed phase (see Fig. 2).
It should be emphasized that the energy density is continuous
at the boundary of the mixed phase and the gaseous phase, hence
the sharpness of the
peak in $c_{\rho}$ is entirely due to the strong temperature
dependence
of $\lambda(T)$ near this boundary. Therefore,
in contrast to the expectation in Refs. \cite{Gu:98,Gu:99}, 
the maximum value of $c_{\rho}$ remains finite
and the peak width in $c_{\rho}(T)$ is not zero in the thermodynamic
limit considered in our study. 

%
%

\vspace*{-0.5cm}

\begin{figure}
\mbox{\psfig{figure=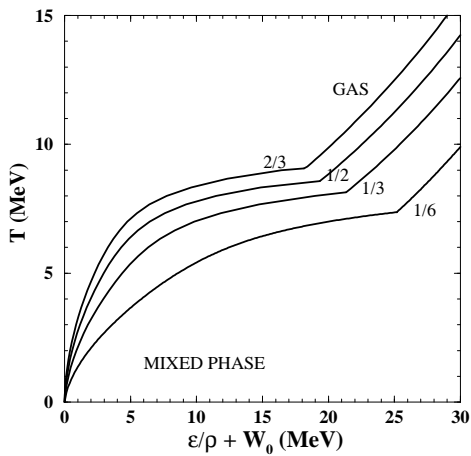,width=98mm}}

\vspace*{-0.5cm}

\caption{\label{fig:three}
Temperature as a function of energy density per nucleon 
(caloric curve)
is shown for fixed nucleon densities ${\rho}/{\rho_{\rm o}} = 1/6, 1/3,
1/2, 2/3$.
}
\end{figure}
%
%

\vspace*{-1.5cm}

\begin{figure}
\mbox{\psfig{figure=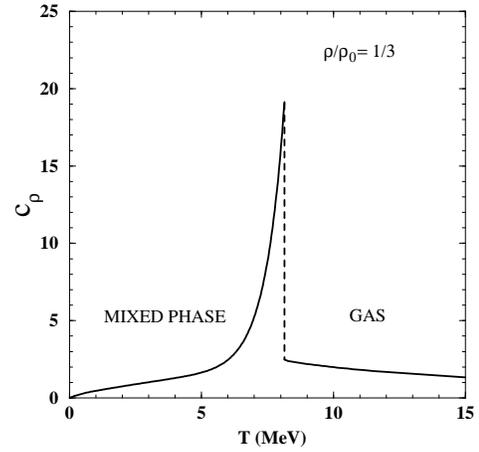,width=98mm}}

\vspace*{-0.5cm}

\caption{\label{fig:four}
Specific heat per nucleon as a function of temperature
at fixed nucleon density ${\rho}/{\rho_{\rm o}} = 1/3$. The dashed line
shows the finite discontinuity of $c_{\rho}(T)$  
at the boundary of the mixed and gaseous phases.
}
\end{figure}


In conclusion, the simplified version of the SMM
is solved analytically
in the grand canonical ensemble.
The progress is achieved by 
reducing the description of phase transitions
to the investigation  of the isobaric
partition function singularities. The model clearly 
demonstrates a 1-st order
phase transition of the liquid-gas type.
The considered system has  peculiar
properties near the boundary of the mixed and gaseous
phases seen in the behavior of the specific heat.

\vspace{0.2cm}
{\bf  Acknowledgments.}  
The authors 
are thankful to 
A.S.~Botvina, Ph.~Chomaz, D.H.E.~Gross, A.D.~Jackson and
J.~Randrup for useful discussions. 
We are thankful to the Alexander von Humboldt Foundation
and DFG, Germany for the financial support. 
The research described in this publication was made possible in part by
Award No. UP1-2119 of the U.S. Civilian Research \& Development
Foundation for the Independent States of the Former Soviet Union
(CRDF).


\end{multicols}
\end{document}